\documentclass[10pt,english]{article}
\usepackage[T1]{fontenc}
\usepackage[latin9]{inputenc}
\usepackage{amsmath}
\usepackage{amssymb}
\usepackage{esint}
\usepackage{feyn}

\makeatletter

\providecommand{\tabularnewline}{\\}

\usepackage{amsfonts} 
\newcommand{\eq}{\begin{equation}}
\newcommand{\eqn}[1]{\label{#1}\end{equation}}
\newcommand{\eea}{\end{eqnarray}}
\newcommand{\eqa}{\begin{eqnarray}}
\newcommand{\eqan}[1]{\label{#1}\end{eqnarray}}
\newcommand{\ba}{\begin{array}}
\newcommand{\ea}{\end{array}}
\newcommand{\eqac}{\begin{equation}\begin{array}{rcl}}
\newcommand{\eqacn}[1]{\end{array}\label{#1}\end{equation}}

\usepackage{babel}

\usepackage{babel}

\usepackage{babel}

\usepackage{babel}

\usepackage{babel}

\usepackage{babel}

\usepackage{babel}

\makeatother

\usepackage{babel}

\makeatother

\usepackage{babel}
\begin{document}

\title{\textbf{N=1 Gribov superfield extension.}\\
 \textbf{ }}

\author{\textbf{M.~M.~Amaral}%
\thanks{macielamaral@uerj.br%
}\,\,,\textbf{ Y.~E.~Chifarelli}%
\thanks{yveseduardo@hotmail.com%
}\,\,, \textbf{V.~E.~R.~Lemes}%
\thanks{vitor@dft.if.uerj.br%
}\,\,\\[2mm] \textit{\small{{{{UERJ $-$ Universidade do Estado
do Rio de Janeiro}}}}}{\small{{{}}}}\\
 {\small{{{ {} }}}}\textit{\small{{{{Instituto de Física $-$
Departamento de Física Teórica}}}}}{\small{{{}}}}\\
 {\small{{{ {} }}}}\textit{\small{{{{Rua S{ã}o Francisco
Xavier 524, 20550-013 Maracan{ã}, Rio de Janeiro, RJ, Brasil.}}}}}}

\maketitle
\vspace{-1cm}

\begin{abstract}

We propose a mechanism displaying confinement, as defined by the behavior
of the propagators, for 4 dimensional, N = 1 supersymmetric Yang-Mills
theory in superfield formalism. In this work we intend to verify the
possibility of extending the known Gribov problem of quantization
of Yang-Mills theories and the implementation of a local action with
auxiliary superfields like Gribov-Zwanziger approach to this problem.

\end{abstract}
\setcounter{page}{0}\thispagestyle{empty}

\vfill{}
 \newpage{}\ \makeatother

\section{Introduction}

The problem of gluon confinement is a challenging issue that is at
the core of the general investigations of strongly coupled gauge theories.

Recently this problem has received great attention in different approaches.
One of these approaches comes from lattice simulation where the behavior
of the gluon propagator in the infrared regime is studied \cite{attilio,tereza,muller}.
These results display positivity violation thus making impossible
a particle interpretation for the gluon excitation at low energies.
This is taken as a strong signal of gluon confinement. In the analytical
point of view, one possible approach of the confinement problem on
Yang-Mills theories (YM), comes from the analysis of the Gribov copies
\cite{gribov}, known as Gribov problem, where the Gribov-Zwanziger
(GZ) model \cite{zwanziger1,zwanziger2,zwanziger25,zwanziger3}, and
this refined version, the so-called Refined Gribov-Zwanziger (RGZ)
model \cite{rgzmodel}, take place. Also, a recently developed model
based on the introduction of a replica of the Faddeev-Popov action
enjoys a confined gluon propagator (replica model) \cite{replicamodel}.
Usually, these models provide propagators behaving as:

\begin{equation}
\mathcal{G}(p^{2})=\frac{p^{2}}{p^{4}+\gamma^{4}},\label{eq:propagatorlikegribov}
\end{equation}
 where $\mathcal{G}(p^{2})$ is the gluon form factor in Euclidean
spacetime and $\gamma$ is a mass parameter (In the GZ model this
parameter is known as the Gribov parameter, which is directly associated
with the restriction of the Feynman path integrals to the Gribov region).

Although it is expected that this problem also occurs in supersymmetric
theories \cite{vanBaalGribovCopies,vanBaalWittenIndex,CapriSorellaSYMcomponentgribov},
there is still a lack of studies on its implementation and consequences.
One of the most interesting possible consequences of that propagator
is that it could solve the problem with infrared singularity of the
first component of the gauge superfield.

In the confining behavior of supersymmetric theories, N = 1, much
has been done since Seiberg's work with super QCD\cite{Seiberg}.
We refere to \cite{terningmodersusy} and references therein that
covers recent developments with nonperturbative results and more.
However here we focus on pure Super Yang-Mills theory (SYM) without
matter.

So in this paper we investigate the SYM (N = 1, D = 4) with superfields
formalism \cite{siegelsuperspace} addressing the Gribov problem as
well as GZ type action. The aim is to investigate how the superfield
extension of these approaches can generate propagators of the confining
type (\ref{eq:propagatorlikegribov}), and thus shed some light on
how the quantization of gauge sector can affect the fermions, even
in non-supersymmetric theories. With the proviso that in the SYM theory
that we study here the fermions are on the adjoint representation
of SU(N), differently from quarks in QCD for example.

The paper is organized as follows. In Sec. 2, we present the euclidean
SYM theory and investigate some details related to gauge-fixing. In
Sec. 3, we present an gauge-invariant local action with auxiliary
(anti)chiral superfields of inspiration in GZ and found that it generates
the desired behavior of the confining SYM propagators. A brief summary
will be devoted in Sec. 4 and notation and useful formulas are presented
in the Appendix.

\section{Superfield approach to Gribov problem, N = 1, D = 4, SYM theory.}

\subsection{N = 1, D = 4, Euclidean SYM theory}

The pure N = 1 Euclidean SYM action on superspace is S' = $S_{SYM}$
+ $\overline{S}_{SYM}$, where $\overline{S}_{SYM}$ is the Osterwalder-Schrader
(OS) conjugate of $S_{SYM}$ \cite{lukierski,morrissuperinstanton,osterwalderschrader}.
We point out here that the Euclidean supersymmetry has its own peculiarities
\cite{Wetterich}. For example, note that hermitian conjugation is
replaced by OS conjugation. Without this we have to work with complex
fields or N \textgreater{} 1. For more details refer the references
above. Keeping this in mind we can work with $S_{SYM}$ \cite{grisarusupergraphs,siegelsuperspace}
given by:

\begin{equation}
S_{SYM}=\frac{1}{64g^{2}}tr\int d^{4}xd^{2}\theta W^{\alpha}W_{\alpha}.\label{eq:actionsym}
\end{equation}
The field strength is given by:

\begin{equation}
W_{\alpha}=\overline{D}^{2}(e^{-gV}D_{\alpha}e^{gV}).
\end{equation}
and covariant derivatives:

\begin{equation}
D_{\alpha}=\frac{\partial}{\partial\theta^{\alpha}}+i\sigma_{\alpha\dot{\alpha}}^{\mu}\bar{\theta}^{\dot{\alpha}}\partial_{\mu}
\end{equation}

\begin{equation}
\bar{D}_{\dot{\alpha}}=-\frac{\partial}{\partial\bar{\theta}^{\dot{\alpha}}}-i\theta^{\alpha}\sigma_{\alpha\dot{\alpha}}^{\mu}\partial_{\mu}.
\end{equation}
The supermultiplet of gauge fields is given by the components of the
superfield $V=V^{a}T_{a}$ (with $V$ real):

\begin{align}
V(x,\theta,\overline{\theta}) & =N(x)+\theta\chi(x)+\bar{\theta}\bar{\chi}(x)+\frac{1}{2}\theta^{2}M(x)+\frac{1}{2}\bar{\theta}^{2}\bar{M}(x)+\nonumber \\
 & \theta\sigma^{\mu}\bar{\theta}a_{\mu}(x)+\frac{1}{2}\bar{\theta}^{2}\theta\lambda(x)+\frac{1}{2}\theta^{2}\bar{\theta}\bar{\lambda}(x)+\frac{1}{4}\theta^{2}\bar{\theta}^{2}D(x).
\end{align}
They belong to the adjoint representation of the gauge group SU(N).
With $[T_{a},T_{b}]=if_{abc}T_{c}$. The gauge transformations are
implicitly defined by:

\begin{equation}
e^{gV^{'}}=e^{ig\bar{\Lambda}}e^{gV}e^{-ig\Lambda}.\label{eq:gaugeexponencial}
\end{equation}
Or, for infinitesimal $\Lambda$:

\begin{equation}
\delta_{gauge}=-\frac{i}{2}L_{gV}(\Lambda+\bar{\Lambda})-\frac{i}{2}(L_{gV}coth(L_{gV/2}))(\Lambda-\bar{\Lambda}),\label{eq:gaugeinfinitesimal}
\end{equation}
with the Lie derivative $L_{gV}X=[gV,X]$ and $\Lambda=\Lambda{}^{a}T_{a}$
a chiral superfield ($\bar{D}_{\dot{\alpha}}\Lambda=0$).

See Appendix for notation and conventions.

\subsection{Gauge-fixing}

Being the action (\ref{eq:actionsym}) gauge invariant, its quantization
is similar to YM theories. Therefore, the functional integration must
be restricted to gauge inequivalente subset of fields and the operator
which appears in the bilinear term is not invertible over the space
of all field configurations so that the propagator necessary to make
the perturbative theory can not be defined unless the set of fields
is restricted. In this case we have to fix the gauge and we can do
covariantly using the usual procedure of Faddeev-Popov (FP) \cite{siegelsuperspace,West}.

Here we present some details of this derivation to show that the Gribov
problem present in YM theories (see \cite{SobreiroSorella,reviewZwanziger}
for a pedagogical reviews) also arises in the generalization of Landau
gauge and FP quantization to SYM theories.

Thus, consider the functional integral for the real scalar gauge superfield
V:

\begin{equation}
Z=\int DVe^{-S_{SYM}(V)},\label{eq:funcional}
\end{equation}
with $S_{SYM}$ given by (\ref{eq:actionsym}) and invariant over
gauge transformations (\ref{eq:gaugeinfinitesimal}). Note that the
operator that appears in the bilinear term is $\partial^{2}\Pi_{\frac{1}{2}}$,
with the spin $\frac{1}{2}$ projection operator given by (\ref{eq:operadorprojetorpimeio}),
is not invertible since it annihilates the chiral (antichiral) superspin
zero parts of V ($\Pi_{0}V=\frac{1}{16\partial^{2}}(\overline{D}^{2}D^{2}+D^{2}\overline{D}^{2})V$).
As a result, the operator has zero modes problems. And we can verify
that they are related to the gauge transformation. Considering a gauge
transformation (\ref{eq:gaugeinfinitesimal}) for $V=0$:

\begin{equation}
\delta_{gauge}=\frac{i}{2}(\bar{\Lambda}-\Lambda),\label{eq:gaugeinfinitesimalVigualzero}
\end{equation}
we found that appear a set of chiral and antichiral fields as in the
case of superspin zero parts of V. Therefore we are integrating over
gauge equivalent fields. As these give rise to zero modes, we are
taking too many configurations into account. We have to choose gauge-fixing
functions corresponding to the chiral (antichiral) gauge parameter
$\Lambda$ that can be taken away by an appropriate gauge transformation.

So we can go ahead with the procedure of FP quantization, where is
inserted into the functional (\ref{eq:funcional}) the identity (which
also defines the FP determinant):

\begin{equation}
\triangle_{F}(V)\int D\Lambda D\bar{\Lambda}\delta(F(V^{\Lambda})-f)\delta(\bar{F}(V^{\Lambda})-\bar{f}),\label{eq:FPdeterminante}
\end{equation}
for any chiral (antichiral) $f$ ($\bar{f}$) and gauge transformations
$F(V)\rightarrow F(V^{\Lambda})$.

As the gauge fixing will be a supersymmetric extension of the Lorentz
(or Landau) gauge $\partial^{\mu}a_{\mu}=0$ and noting that $\partial^{\mu}a_{\mu}$
is a component of chiral (antichiral) superfield $\overline{D}^{2}D^{2}V$
($D^{2}\overline{D}^{2}V$), we must implement the conditions $\overline{D}^{2}D^{2}V=0$,
$D^{2}\overline{D}^{2}V=0$. Therefore in the quantization procedure,
$F=\overline{D}^{2}D^{2}V$ and $\bar{F}=D^{2}\overline{D}^{2}V$
are a suitable gauge-fixing functions. 

And following the usual procedure we ended with the action of gauge
fixing (Landau gauge)

\begin{equation}
S_{gf}=-\frac{1}{16}s\{tr\int d^{4}xd^{4}\theta(c'D^{2}V+\bar{c}'\overline{D}^{2}V)\},\label{eq:actiongaugefix}
\end{equation}
where the Faddeev-Popov ghost fields will be chiral (antichiral) superfield
as the gauge parameter $\Lambda$ ($\bar{\Lambda}$).

$c'=c'^{a}T_{a}$ and $c=c^{a}T_{a}$ are the antighost and the ghost
respectively. And $s$ is the BRST nilpotent operator ($s^{2}=0)$.

The total action $S_{SYM}+S_{gf}$ is invariant under the BRST transformations
\cite{siegelsuperspace}:

\[
sV=\delta_{\Lambda}V\mid_{\Lambda=ic}
\]

\begin{align}
sV & =[\frac{1}{2}L_{gV}(c+\bar{c})+\frac{1}{2}(L_{gV}coth(L_{gV}))(c-\bar{c})]\nonumber \\
 & =\{-\bar{c}+c+\frac{1}{2}[gV,c+\bar{c}]+...\}\nonumber \\
sc & =-c^{2}(sc^{a}=\frac{i}{2}f_{abc}c^{b}c^{c})\nonumber \\
s\bar{c} & =-\bar{c}^{2}\nonumber \\
sc' & =B\nonumber \\
s\bar{c}' & =\bar{B}\nonumber \\
sB & =0\nonumber \\
s\bar{B} & =0.\label{eq:brst}
\end{align}

With $s$ carrying ghost number 1.

The ghost part of gauge fixing action becomes:

\begin{equation}
S_{FP}=\frac{1}{16}tr\int d^{4}xd^{4}\theta(c'D^{2}sV+\bar{c}'\overline{D}^{2}sV).\label{eq:actionghost}
\end{equation}

An important detail is that the Jacobian $\triangle_{F}(V)$ in (\ref{eq:FPdeterminante})
contains a determinant of type FP:

\[
(\frac{\delta F(V^{\Lambda})}{\delta\Lambda}+\frac{\delta F(V^{\Lambda})}{\delta\bar{\Lambda}})(\frac{\delta\bar{F}(V^{\Lambda})}{\delta\Lambda}+\frac{\delta\bar{F}(V^{\Lambda})}{\delta\bar{\Lambda}}).
\]

With the variational derivatives of $F$ and $\bar{F}$ evaluate at
$\bar{\Lambda}=\Lambda=0$. And so in our case we have the operators:
\begin{equation}
\overline{D}^{2}D^{2}[-\frac{i}{2}L_{gV}\bullet-\frac{i}{2}(L_{gV}coth(L_{gV/2}))\bullet]\label{eq:operadorgeneralizaFP}
\end{equation}

\begin{equation}
D^{2}\overline{D}^{2}[-\frac{i}{2}L_{gV}\bullet-\frac{i}{2}(L_{gV}coth(L_{gV/2}))\bullet],\label{eq:operadorgeneralizaFP-1}
\end{equation}
where the symbol $\bullet$ is to indicate that those operators act
in chiral and antichiral superfields, which may have zero mode problems
making the gauge-fixing still incomplete.

\subsection{Gribov problem}

Similarly to YM let us now explicitly show that in the Landau gauge
the gauge condition is not ideal as already noted above. Consider
two equivalents fields $V$ e $V'$ connected by the gauge transformation
(\ref{eq:gaugeinfinitesimal}), if both satisfy the same condiction
of Landau gauge $\overline{D}^{2}D^{2}V=0$, $D^{2}\overline{D}^{2}V=0$
e $\overline{D}^{2}D^{2}V'=0$, $D^{2}\overline{D}^{2}V'=0$, we have

\begin{equation}
\overline{D}^{2}D^{2}((\bar{\Lambda}-\Lambda)-\frac{1}{2}[gV,\bar{\Lambda}+\Lambda]+...)=0\label{eq:superlandauop}
\end{equation}

\begin{equation}
D^{2}\overline{D}^{2}((\bar{\Lambda}-\Lambda)-\frac{1}{2}[gV,\bar{\Lambda}+\Lambda]+...)=0.\label{eq:superlandauop-1}
\end{equation}

And onshell, the problem of zero modes become ($\overline{D}^{2}D^{2}\Lambda=16\partial^{2}\Lambda$
from (\ref{eq:derivativecomutator})):

\begin{equation}
(-16\partial^{2}\bullet-\frac{1}{2}[gV,16\partial^{2}\bullet]+...)\Lambda=0\label{eq:superlandauop-2}
\end{equation}

\begin{equation}
(16\partial^{2}\bullet-\frac{1}{2}[gV,16\partial^{2}\bullet]+...)\bar{\Lambda}=0.\label{eq:superlandauop-1-1}
\end{equation}

Thus the existence of infinitesimal copies even after FP quantization
is linked to the existence of zero modes of the operators above, that
we can see as the supersymmetric generalization of FP operator.

For zero V we can construct in the space of chiral and antichiral
superfields the eigenvalue equations

\begin{equation}
-16\partial^{2}\Lambda=\lambda_{1}\Lambda\label{eq:superlandauop-1-1-1-1}
\end{equation}

\begin{equation}
16\partial^{2}\bar{\Lambda}=-\lambda_{2}\bar{\Lambda}\label{eq:superlandauop-1-1-1-1-1}
\end{equation}
that only has positive eigenvalues $\lambda_{1}=\lambda_{2}=16p^{2}>0$.
Thus for small values of V, we can expect that the eigenvalues $\lambda_{1}(V)$
e $\lambda_{2}(V)$ are greater than zero. However, for large V, it
can no longer be guaranteed, and may appear negative eigenvalues of
V sufficiently large. And thus the above operators will also have
null eigenvalues. This means that our gauge condition is not ideal.
It is noteworthy that chiral and antichiral eigenvalues equations
are present, e.g., in context of superinstantons \cite{morrissuperinstanton,doreymanyinstanton,shifmaninstantonsusy}.

To view the issue of zero modes let us take the equation (\ref{eq:superlandauop-1-1-1-1})
as an equation of eigenvalues to first order in V:

\begin{equation}
(-16\partial^{2}\bullet-\frac{1}{2}[gV,16\partial^{2}\bullet])\Lambda=\lambda\Lambda.\label{eq:superlandauop-2-1}
\end{equation}

We remark that as in the non supersymmetric case we can understand
this equation as an Schrondinger type equation (being here supersymmetric),
where the second member plays the role of a potential, and study its
eigenvalues. For example see reference \cite{hendriksuperschoreq}
with a delta potential and references therein.

Therefore, we want to show that when generalize the Landau gauge and
FP quantization from YM to SYM with superfields in N = 1 and D = 4,
the well-known Gribov problem can also be extended. Remembering that
we have defined the Gribov problem through the coupled eigenvalue
equations and in the non supersymmetric case zero modes are well know
for Gribov problem, without going into more details in this paper,
we will see in the next section how to implement a possible solution,
the local action of super GZ, and how it modifies the propagators
for the confining type (\ref{eq:propagatorlikegribov}).

\section{Gribov-Zwanziger local action on superspace}

Now that we have shown that the FP quantization is incomplete in the
sense of the previous section, we need to improve the gauge fixing.
Gribov propose for YM theories to further restrict to a region of
integration, the so-called Gribov region, which is defined as the
region of gauge fields obeying the Landau gauge and for which the
Faddeev-Popov operator is positive definite. Whereas in our case the
operator is as it appears in (\ref{eq:superlandauop-2}) and (\ref{eq:superlandauop-1-1}).
We then assume that this procedure can be generalized and thus the
region of integration can be restrict in superspace.

To implement such a restriction we are going directly to the GZ approach
to construct an action which implements the restriction to the Gribov
region order by order. We begin by observing that the operators in
question are (within a factor $\pm i$) the same that appear in the
ghosts sector (\ref{eq:actionghost}), i.e., the expectation value
of the Faddeev\textendash{} Popov operator is the inverse ghost propagator,
which suggests how the auxiliary fields Zwanziger-style must be entered
in the action. This is done by introducing auxiliary superfields in
the form of two quartets of BRS, one with chiral:

\begin{align}
sw'=u', & \; su=w\nonumber \\
su'=0, & \; sw=0,\label{eq:brst-1}
\end{align}
and another with antichiral superfields:

\begin{align}
s\bar{w}' & =\bar{u}',\; s\bar{u}=\bar{w}\nonumber \\
s\bar{u}' & =0,\; s\bar{w}=0,\label{eq:brst-1-1}
\end{align}

At this point is important for our construction to show the ultraviolet
dimension and ghost number of all fields and operators.

\begin{table}[h]
 \centering %
\begin{tabular}{|c|c|c|c|c|c|c|c|c|c|c|}
\hline 
fields and operators  & $\theta^{\alpha}$  & $D_{\alpha}$  & $V$  & $c'$  & $c$  & $B$  & $w'$  & $w$  & $u'$  & $u$ \tabularnewline
\hline 
UV dimension  & -$\frac{1}{2}$  & $\frac{1}{2}$  & 0  & 1  & 0  & 1  & 1  & 0  & 1  & 0 \tabularnewline
\hline 
Ghost number  & 0  & 0  & 0  & -1  & 1  & 0  & -1  & 1  & -1  & 1 \tabularnewline
\hline 
\end{tabular}\caption{Quantum numbers of fields and operators.}

\label{table1} 
\end{table}

Thus, keeping in mind the non-supersymmetric approach, we propose
the super GZ action:

\begin{align}
S_{SGZ} & =tr\int d^{4}xd^{4}\theta s\{w'D^{2}[\frac{1}{2}L_{gV}(u+\bar{u})+\frac{1}{2}(L_{gV}coth(L_{gV}))(u-\bar{u})]+\nonumber \\
 & \bar{w}'\overline{D}^{2}[\frac{1}{2}L_{gV}(u+\bar{u})+\frac{1}{2}(L_{gV}coth(L_{gV}))(u-\bar{u})]\}+\nonumber \\
 & \gamma^{2}tr\int d^{4}xd^{4}\theta V(u-\bar{u})+\nonumber \\
 & \gamma^{2}tr\int d^{4}xd^{2}\theta Vu'+\nonumber \\
 & \gamma^{2}tr\int d^{4}xd^{2}\bar{\theta}V\bar{u}'.\label{eq:supergzaction}
\end{align}

Where $\gamma^{2}$ is a mass parameter, which should be determined
by the theory, shown below, must be nonzero.

Thus, the total action is: 
\begin{equation}
S=S_{SYM}+S_{gf}+S_{SGZ}.\label{eq:totalaction}
\end{equation}

At this point is important to remember that the term $\gamma^{2}tr\int d^{4}xd^{4}\theta V(u-\bar{u})$
breaks the BRST symmetry as in the non supersymmetric GZ action. This
term is a soft breaking term and there are many methods in order to
ensure the ultraviolet renormalization of the action. From the introduction
of classical sources in order to treat this soft breaking term as
an insertion to a mechanism that transform this breaking into a classical
linear breaking. It is not the purpose of this work the study of the
renormalizability of the action. In spite of that the way in order
to generalise the GZ action to a supersymmetric one is so close to
the original GZ procedure that we expect that the supersymmetric GZ
also is ultraviolet renormalizable.

With GZ action generalization at our disposal we can now calculate
the propagators and ensure they have the expected behavior that occurs
in confining YM theories and analyze other features it adds to SYM.

\subsection{The super gauge propagator}

First we calculate the propagator for gauge superfield V. To calculate
the gauge propagator we need only the bilinear of S. Thus, for $S_{SGZ}$,
we have:

\begin{align}
S_{SGZ2} & =tr\int d^{4}xd^{4}\theta s\{w'D^{2}(u-\bar{u})+\nonumber \\
 & \bar{w}'\overline{D}^{2}(u-\bar{u})\}+\nonumber \\
 & \gamma^{2}tr\int d^{4}xd^{4}\theta V(u-\bar{u})+\nonumber \\
 & \gamma^{2}tr\int d^{4}xd^{2}\theta Vu'+\nonumber \\
 & \gamma^{2}tr\int d^{4}xd^{2}\bar{\theta}V\bar{u}',\label{eq:supergzaction-1}
\end{align}
and for terms with $u,u'$:

\begin{align}
S_{SGZ2} & =tr\int d^{4}xd^{2}\theta(-\frac{1}{4}u'\overline{D}^{2}D^{2}u+\frac{1}{4}\gamma^{2}V\overline{D}^{2}\bar{u}+\gamma^{2}Vu')+\nonumber \\
 & tr\int d^{4}xd^{2}\bar{\theta}(+\frac{1}{4}\bar{u}'D^{2}\overline{D}^{2}\bar{u}-\frac{1}{4}\gamma^{2}VD^{2}u+\gamma^{2}V\bar{u}').\label{eq:supergzaction-1-1}
\end{align}
With give one contribuition to bilinear term ($\overline{D}^{2}D^{2}u=16\partial^{2}u)$

\begin{align}
S_{SGZ2} & =-tr\int d^{4}xd^{4}\theta V\frac{\gamma^{4}}{2\partial^{2}}V.\label{eq:supergzaction-1-1-1}
\end{align}
So, the free total action yields the field equation for V, inserting
your source $J_{V}$:

\begin{equation}
\frac{1}{16}D^{\alpha}\overline{D}^{2}D_{\alpha}V-\frac{\gamma^{4}}{2\partial^{2}}V+\frac{1}{16}D^{2}B+\frac{1}{16}\overline{D}^{2}\overline{B}=-J_{V}.
\end{equation}
Using transverse operator $\Pi_{\frac{1}{2}}$, and $\Pi_{\frac{1}{2}}V=V$,
this action give rise to a propagator of the form (in space coordinates):

\begin{equation}
\triangle_{VV}^{c}(1,2)=-\frac{2\partial^{2}}{\partial^{4}+\gamma^{4}}\Pi_{\frac{1}{2}}\delta^{4}(x_{1}-x_{2})\mbox{\ensuremath{\delta^{4}}(\ensuremath{\theta_{1}}-\ensuremath{\theta_{2}})}.
\end{equation}
Where $\delta^{4}(\theta_{1}-\theta_{2})=(\theta_{1}-\theta_{2})^{2}(\bar{\theta_{1}}-\bar{\theta_{2}})^{2}$.

To see how the introduction of $S_{SGZ}$ brings light on confinement
of both bosons as fermions, we shall observe the propagators in field
components.

Using (\ref{eq:projtransondeltaV}) we can project the propagator
for the gauge field $a_{\mu}$ and gaugino $\lambda^{\alpha}$:

\begin{equation}
\triangle_{a_{\mu}a_{\nu}}^{c}(1,2)=-\frac{2\partial^{2}}{\partial^{4}+\gamma^{4}}(\delta_{\mu\nu}-\frac{2\partial_{\mu}\partial_{\nu}}{\partial^{2}})\delta^{4}(x_{1}-x_{2})
\end{equation}

\begin{equation}
\triangle_{\lambda\bar{\lambda}}^{c}(1,2)=\frac{5}{2}\frac{i\partial^{2}}{\partial^{4}+\gamma^{4}}\sigma^{\mu}\partial_{\mu}\delta^{4}(x_{1}-x_{2}).
\end{equation}

And we found that both show behavior as occurs for gluons in non-supersymmetric
theories (\ref{eq:propagatorlikegribov}). Namely, we obtain propagators
with confining behavior in an integrated manner for bosons and fermions
in this model.

Another field component projection we analyze is the dimensionless
and massless component of the gauge superfield $V$ (the $\theta=0$
component) which propagator also becomes Gribov type 

\begin{equation}
\triangle_{NN}^{c}(1,2)=\frac{4}{\partial^{4}+\gamma^{4}}\delta^{4}(x_{1}-x_{2}).
\end{equation}

And so solves naturally a problem characteristic of supersymetrics
theories in four dimensions, that is the appearance of a infrared
singularity in this $V$ component \cite{piguetnotasaula}, here named
$N(x)$. This also indicates that the parameter $\gamma$ must be
different from zero, at least in this framework, to avoid this infrared
singularity.

\subsection{Ghost propagators and $\gamma$ parameter}

Since the action (\ref{eq:totalaction}) only makes sense if the $\gamma$
parameter is nonzero, we will now explicitly show that it is not independent
in this theory. Its determination is closely linked to the restriction
of the functional integration to the first Gribov region, which we
will discuss some details here.

First, it is noteworthy that in the literature dealing with the Gribov
problem in YM theories there are recent consensus on the scenario
of dominance of configurations on the Gribov horizon on the Landau
gauge \cite{reviewZwanziger}, so that the restriction to the first
Gribov region is, in practice, to take the configurations on the horizon,
ie where occur the zeros modes of the FP operator, in our case given
by equations (\ref{eq:operadorgeneralizaFP}, \ref{eq:operadorgeneralizaFP-1}).
Second, and as we have pointed out in the introduction of super GZ,
calculate the propagator of the ghosts is to take the inverse of these
operators. So we focus on these calculus to one loop order to establish
the one loop gap equation Gribov style.

In order to characterize the integration in the first Gribov region
it is important to remember that the two point ghost function is essentially
the inverse of the Faddev-Popov operator and the zero eigenvalue of
the Gribov equation corresponds to a exactly to the Gribov frontier.
In these sense the two point ghost function goes to infinity at the
Gribov frontier. These condition is the most simpler way to obtain
the gap equation for $\gamma$. These procedure is explained in details
in \cite{gribov} and is easily extended to the $N=1$ supersymmetric
case. First we need to calculate the two point ghost function, using
perturbation theory these is, at first order of the form:

\begin{center}
\begin{tabular}{ccc}
 &  & \tabularnewline
$\Diagram{\\
\vertexlabel_{c'_{a}}h & \momentum{hA}{p} & h\vertexlabel_{c{}_{b}}
}
$  & +  & $\Diagram{ &  & \momentum{glA}{k}\\
\vertexlabel_{c'_{a}}\momentum{hA}{p} & h & \momentum{hA}{p-k} & \momentum{hA}{p}\vertexlabel_{c{}_{b}}
}
$\tabularnewline
 &  & \tabularnewline
\end{tabular}
\par\end{center}

Where the line between $c'_{a}$ and $c_{b}$ corresponds to the zero
order super ghost propagators $\mathcal{G}_{c'c}^{ab0}=-\triangle_{c'c}^{c}(1,2)$.
After a straightforward calculation:

\begin{equation}
\triangle_{c'c}^{c}(1,2)=\frac{1}{\partial^{2}}\bar{D}^{2}\delta^{4}(x_{1}-x_{2})\mbox{\ensuremath{\delta^{4}}(\ensuremath{\theta_{1}}-\ensuremath{\theta_{2}})}\label{eq:ghostpropagator1}
\end{equation}

\begin{equation}
\triangle_{\bar{c}'\bar{c}}^{c}(1,2)=-\frac{1}{\partial^{2}}D^{2}\delta^{4}(x_{1}-x_{2})\mbox{\ensuremath{\delta^{4}}(\ensuremath{\theta_{1}}-\ensuremath{\theta_{2}}).}
\end{equation}
And we can define in momentum space, the one loop corrected ghost
propagator as

\begin{equation}
\mathcal{G}_{c'c}^{ab}=(\mathcal{G}_{c'c}^{ab0}+\mathcal{G}_{c'c}^{ab1}),\label{eq:ghostpropzeroplusoneorder}
\end{equation}
according to diagram above. With $\mathcal{G}_{c'c}^{ab0}$ given
from (\ref{eq:ghostpropagator1}):

\begin{equation}
\mathcal{G}_{g}^{ab0}=-\frac{\delta^{ab}}{p^{2}}\bar{D}^{2}\delta^{4}(\theta_{1}-\theta_{2}).
\end{equation}
Using the improved Feynman rules (and D algebra) from \cite{grisarusupergraphs,siegelsuperspace,Srivastava}
and after delta functions and D derivatives manipulations, we have:

\begin{align}
\mathcal{G}_{c'c}^{ab1} & =-(2\pi)^{4}g^{2}\delta^{ab}\frac{1}{p^{2}}\bar{D}_{1}^{2}(p)\delta^{4}(\theta_{1}-\theta_{2})\int\frac{d^{D}k}{(2\pi)^{D}}\frac{k^{2}}{k^{4}+\gamma^{4}}\frac{1}{(p-k)^{2}}.
\end{align}
Next we define:

\begin{align}
\sigma(p^{2},\gamma^{2}) & =(2\pi)^{4}g^{2}\int\frac{d^{D}k}{(2\pi)^{D}}\frac{k^{2}}{k^{4}+\gamma^{4}}\frac{1}{(p-k)^{2}}.
\end{align}
Therefore, from (\ref{eq:ghostpropzeroplusoneorder}):

\begin{align}
\mathcal{G}_{c'c}^{ab} & =-\delta^{ab}\frac{1}{p^{2}}\bar{D}_{1}^{2}\delta^{4}(\theta_{1}-\theta_{2})(1+\sigma).
\end{align}
Resumming the one-particle reducible diagrams gives:

\begin{align}
\mathcal{G}_{c'c}^{ab} & =-\delta^{ab}\frac{1}{p^{2}}\bar{D}_{1}^{2}\delta^{4}(\theta_{1}-\theta_{2})\frac{1}{(1-\sigma)}.
\end{align}
This means that, as with Gribov problem in YM theories, the super
ghost propagator is enhanced.

Now, as we are interested in the low momentum behavior we analyze
the behavior of $(1-\sigma)$ with $k\approx0$, ie we get $\sigma(0,\gamma^{2})$:

\begin{align}
\sigma(0,\gamma^{2}) & =(2\pi)^{4}g^{2}\int\frac{d^{D}k}{(2\pi)^{D}}\frac{1}{k^{4}+\gamma^{4}}.
\end{align}
And so we are able to define the one loop gap equation according to
the above discussion of the scenario of dominance of configurations
on the Gribov horizon, ie the ghost propagator (the inverse of FP
operator) going to infinity, $(1-\sigma)=0$:

\begin{equation}
(2\pi)^{4}g^{2}\int\frac{d^{D}k}{(2\pi)^{D}}\frac{1}{k^{4}+\gamma^{4}}=1.\label{eq:oneloopgapequation}
\end{equation}
It is worth mentioning that calculating to one loop the propagator
for the antichirais fields $\mathcal{G}_{\bar{c}'\bar{c}}^{ab}$ similarly
we have

\begin{equation}
\mathcal{G}_{\bar{c}'\bar{c}}^{ab}=\delta^{ab}\frac{1}{p^{2}}D_{1}^{2}\delta^{4}(\theta_{1}-\theta_{2})\frac{1}{(1-\sigma)},
\end{equation}
such that it has the same integral (\ref{eq:oneloopgapequation})
which defines the $\gamma$ parameter.

Note that the integral in equation (\ref{eq:oneloopgapequation})
can be well defined with dimensional regularization and that similar
calculations can be found in the literature, see the already mentioned
review \cite{SobreiroSorella} for example.

Thus the $\gamma$ parameter is not independent, being defined for
the one loop gap equation (\ref{eq:oneloopgapequation}). It is clear
that in close analogy to the Gribov-Zwanziger procedure \cite{zwanziger2,zwanziger25,zwanziger3}
it is also possible to work directly with the gap equation $\frac{\delta\Gamma}{\delta\gamma^{2}}=0$.
The results will be the same as in the more simple method explained
in these section.

\section{Conclusions}

In this work we have studied the super Yang-Mills theory, N = 1, D
= 4, considering the Gribov problem present in YM theories as well
as a generalization of the Gribov-Zwanziger approach with auxiliary
superfields in superspace. With this approach we find that the confining
behavior is therefore induced in this supersymmetric theory. The results
presented here are a first step toward a more extensive investigation
but suggests further that the Gribov problem and its solutions can
be treated consistently in supersymmetric theories and that this can
shed light on wider issues of these theories. In our case this approach
allows to solve the well known problem of infrared supersymetric theories
in four dimensional space and the theory becomes confining.

It is worth mentioning that we believe to be possible to make a renormalization
of this theory since the inclusion of auxiliary superfields the only
term that break supersymmetry is a soft breaking term. This possibility
is under investigation. Further possibility under investigation is
the possibility that a supersymmetric breaking together with the Gribov
mechanism can be important in the confinement of the fermions. The
construction of a supersymmetric breaking that preserves the confining
propagators for fermions and that leads them on fundamental representation
of SU(N) are the next step to be done in these direction and is also
under study. Other research possibilities in this framework are the
supersymmetric version of RGZ and replica model mentioned in the introduction,
as well as models with extended supersymmetry. Still would be interesting
to check possibles condensates.

Finally we would like to express our hope that the, 4 dimensional
N = 1, superfields formalism can serve as a laboratory to bring some
insigth on non-supersymmetric theories such as QCD where there is
still a search for an integrated treatment for problems such as non-perturbative
confinement of quarks and gluons and chiral symmetry breaking (see
introduction of \cite{Maas} for some discution).

\section*{Acknowledgements}

The Conselho Nacional de Desenvolvimento Científico e tecnológico
CNPq- Brazil, Fundação de Amparo a Pesquisa do Estado do Rio de Janeiro
(Faperj). the SR2-UERJ and the Coordenação de Aperfeiçoamento de Pessoal
de Nível Superior (CAPES) are acknowledged for the financial support. 

\appendix

\section[Appendix]{Notation, conventions and some useful formulas}

We work with Euclidean metric: $\delta\mu\nu$=diag(++++), with Wick
rotation from a theory in Minkowski space: $d^{4}x\rightarrow id^{4}x$,
$x^{4}\rightarrow ix^{0}$. Euclidean $\sigma$ - matrices ($\sigma_{i}$
- pauli matrices) are defined as follows:

\begin{equation}
\sigma^{\mu}=\sigma_{\mu}=(\sigma_{i},i)
\end{equation}

\begin{equation}
\bar{\sigma}^{\mu}=\bar{\sigma}_{\mu}=(-\sigma_{i},i),
\end{equation}
witch are OS self-conjugate, and can include the following relations:

\begin{equation}
\bar{\sigma}^{\mu\alpha\dot{\alpha}}=\varepsilon^{\dot{\alpha}\dot{\beta}}\varepsilon^{\alpha\beta}\sigma_{\beta\dot{\beta}}^{\mu}
\end{equation}

\begin{equation}
tr\sigma_{\mu}\sigma_{\nu}=-2\delta_{\mu\nu}
\end{equation}

\begin{equation}
\sigma_{\dot{\alpha}\alpha}^{\mu}\bar{\sigma}_{\mu}^{\dot{\beta}\beta}=-2\delta_{\alpha}^{\beta}\delta_{\dot{\alpha}}^{\dot{\beta}}.
\end{equation}

Some supersymmetrics conventions and useful formulas:

\begin{equation}
\theta^{\alpha}\theta^{\beta}=-\frac{1}{2}\epsilon^{\alpha\beta}\theta^{2}
\end{equation}

\begin{equation}
\bar{\theta}^{\dot{\alpha}}\bar{\theta}^{\dot{\beta}}=\frac{1}{2}\epsilon^{\dot{\alpha}\dot{\beta}}\bar{\theta}^{2}.
\end{equation}

Covariant derivatives:

\begin{equation}
D_{\alpha}=\frac{\partial}{\partial\theta^{\alpha}}+i\sigma_{\alpha\dot{\alpha}}^{\mu}\bar{\theta}^{\dot{\alpha}}\partial_{\mu}
\end{equation}

\begin{equation}
\bar{D}_{\dot{\alpha}}=-\frac{\partial}{\partial\bar{\theta}^{\dot{\alpha}}}-i\theta^{\alpha}\sigma_{\alpha\dot{\alpha}}^{\mu}\partial_{\mu}
\end{equation}

\begin{equation}
\{D_{\alpha},\bar{D}_{\dot{\alpha}}\}=-2i\sigma_{\alpha\dot{\alpha}}^{\mu}\partial_{\mu}
\end{equation}

\begin{align}
 & [D^{2},\overline{D}^{2}]=-8i(D\sigma^{\mu}\overline{D})\partial_{\mu}-16\partial^{2}\nonumber \\
 & =8i(\overline{D}\bar{\sigma}^{\mu}D)\partial_{\mu}+16\partial^{2}\label{eq:derivativecomutator}
\end{align}

\begin{equation}
\int d^{2}\theta=-\frac{1}{4}D^{2},\int d^{2}\overline{\theta}=-\frac{1}{4}\overline{D}^{2}.
\end{equation}

Note that with these definitions our notation in superspace take the
same form as in \cite{wessbagger}.

Finally we present the projection operators and their relations with
delta functions:

\begin{equation}
\Pi_{\frac{1}{2}}=-\frac{D^{\alpha}\overline{D}^{2}D_{\alpha}}{8\partial^{2}}=-\frac{\overline{D}^{\alpha}D^{2}\overline{D}_{\alpha}}{8\partial^{2}}\label{eq:operadorprojetorpimeio}
\end{equation}

\begin{equation}
\Pi_{0^{+}}=\frac{\overline{D}^{2}D^{2}}{16\partial^{2}}
\end{equation}

\begin{equation}
\Pi_{0^{-}}=\frac{D^{2}\overline{D}^{2}}{16\partial^{2}}
\end{equation}

\begin{equation}
\Pi_{0}=\Pi_{0^{-}}+\Pi_{0^{+}},\Pi_{0}+\Pi_{\frac{1}{2}}=1
\end{equation}

\begin{equation}
\Pi_{\frac{1}{2}}\delta^{4}(\theta_{1}-\theta_{2})=-\frac{1}{2\partial^{2}}e^{i(\theta_{2}\sigma^{\mu}\bar{\theta}_{1}-\theta_{1}\sigma^{\mu}\bar{\theta}_{2})\partial_{\mu}}(4-\partial^{2}(\theta_{1}-\theta_{2})^{2}(\bar{\theta_{1}}-\bar{\theta_{2}})^{2})\label{eq:projtransondeltaV}
\end{equation}

\begin{equation}
\bar{D_{1}}^{2}\mbox{\ensuremath{\delta^{4}}(\ensuremath{\theta_{1}}-\ensuremath{\theta_{2}})}=-4(\theta_{1}-\theta_{2})^{2}e^{i(\theta_{1}\sigma^{\mu}\bar{\theta}_{1}-\theta_{1}\sigma^{\mu}\bar{\theta}_{2})\partial_{\mu}}\label{eq:deltaS}
\end{equation}

\begin{equation}
D_{1}^{2}\mbox{\ensuremath{\delta^{4}}(\ensuremath{\theta_{1}}-\ensuremath{\theta_{2}})}=-4e^{-i(\theta_{1}\sigma^{\mu}\bar{\theta}_{1}-\theta_{2}\sigma^{\mu}\bar{\theta}_{1})\partial_{\mu}}(\bar{\theta}_{1}-\bar{\theta}_{2})^{2}.\label{eq:deltaSbarra}
\end{equation}

\end{document}